\documentclass[pra,twocolumn,showpacs,preprintnumbers,amsmath,linenumbers,amssymb]{revtex4}
\usepackage{amsfonts}
\usepackage{mathrsfs}
\usepackage{amsmath}
\usepackage{amssymb}
\usepackage{txfonts}

\usepackage{graphicx}
\usepackage{dcolumn}
\usepackage{bm}
\usepackage{color}

\begin{document}



\title{High Order Momentum Modes by Resonant Superradiant
Scattering }

\author{Xiaoji Zhou}\thanks{Electronic address: xjzhou@pku.edu.cn }
\affiliation{School of Electronics Engineering $\&$ Computer
Science, Peking University, Beijing 100871, China}
\author{Jiageng Fu}
\affiliation{School of Electronics Engineering $\&$ Computer
Science, Peking University, Beijing  100871, China}
\author{Xuzong Chen}
\affiliation{School of Electronics Engineering $\&$ Computer
Science, Peking University, Beijing  100871, China}


\begin{abstract}
The spatial and time evolutions of superradiant scattering are
studied theoretically for a weak pump beam with different frequency
components traveling along the long axis of an elongated
Bose-Einstein condensate. Resulting from the analysis for mode
competition between the different resonant channels and the local
depletion of the spatial distribution in the superradiant Rayleigh
scattering, a new method of getting a large number of high-order
forward modes by resonant frequency components of the pump beam is
provided, which is beneficial to a lager momentum transfer in atom
manipulation for the atom interferometry and atomic optics.

\end{abstract}


\maketitle
\section{Introduction}

Atom interferometry is a valuable tool for studying scientific and
technical fields such as precision measurements and quantum
information, and a very bright source is Bose-Einstein Condensate
(BEC) of atomic gases. In which an important technique is to
manipulate the translational motion of atoms and transfer atoms
coherently between two localities in position and
momentum~\cite{cronin}. To obtain the momentum transfer, one useful
method is two-photon Bragg diffraction, where two laser beams
impinge upon atoms, whose atoms can undergo stimulated
light-scattering events by absorbing a photon from one of the beams
and emitting into the other. The momentum transfer is determined by
the difference in the wave vectors of the beams, and the frequency
difference defines the corresponding energy
transfer~\cite{Brunello}. We will introduce another method to obtain
a large number of high-order momentum modes by the resonant
superradiant scattering from a BEC for a weak pump beam with several
frequency components.

A typical superradiance experiment consists in a far off-resonant
laser pulse traveling along the short axis of a cigar-shaped BEC
sample~\cite{Inouye1999science}, the scattered lights, called
end-fire modes, propagate along the long axis of the condensate, and
the recoiled atoms are refereed to as side modes.  A series of
experiments ~\cite{Schneble2003scince, 1999, Bar-Gill2007arxiv,
Courteille2,sadler} have sparked related interests in phase-coherent
amplification of matter waves ~\cite{Schneble2003scince,1999},
quantum information ~\cite{Bar-Gill2007arxiv}, collective scattering
instability~\cite{Courteille2}, and coherent imaging~\cite{sadler}.
Several theoretical descriptions of these cooperative scattering in
BEC with single-frequency pump have also been presented
~\cite{Moore1999prl, Pu2003prl, Zobay2006pra,guo}.

For the long and weak pump beam, we can observe the forward peaks
correspond to Bragg diffraction of atoms ~\cite{Inouye1999science},
where the high order scattering is limited by detuning barriers for
the end-fire mode radiation~\cite{Zobay1}. On the other hand, a
X-shaped recoiling pattern is demonstrated in a short and strong
pulse as Kapitza-Dirac diffraction of
atoms~\cite{Schneble2003scince}, where an atom in the condensate
absorbs a photon from the pump laser, then emits a photon into an
end-fire mode, and recoils forwardly. Meanwhile another atom absorbs
a photon from the end-fire modes, emits into the pump beam  and
finally recoils backwardly. In this case, there is an energy
mismatch of four times the one-photon recoil kinetic energy $\hbar
\omega_{r}$ in backward scattering, which then remains very weak
unless a short pumping pulse with a broad spectrum is used. Hence,
two phase-locked incident lasers with the frequency difference
$\Delta \omega$ compensating for the energy mismatch has been used
~\cite{Bar-Gill2007arxiv,yang,Cola}, which is named resonant
superradiance, where a large number of backward recoiling atoms can
be produced.

Followed that, it is attractive to extent this idea to achieve a
high momentum transfer by overcoming the detuning barriers, by a
weak and long pump beams with the resonant frequency. It requires to
analysis the competition between the different transition channels
and the spatial distribution of different modes. Because the above
traditional superradiant-scattering configuration involves many
atomic side modes coupled together, to simplify it, we chose another
configuration where a pump beam travels along the long axis of the
BEC. This scheme is widely studied in photon echo ~\cite{Piovella},
decoherence~\cite{2005Italy}, spatial distribution effects~\cite{li}
and self-organized formation of dynamic gratings~\cite{Hilliard}.
Since the pulse length is far longer than the initial spontaneous
process~\cite{Zobay2006pra}, we choose the semi-classical theory
which can well describe the experimental
results~\cite{Zobay2006pra,Bar-Gill2007arxiv,yang}.

In this paper, we first introduce the semi-classical theory for the
superradiance scattering with a several-frequency pump in the weak
coupling. Then the spatial and time evolutions of scattered modes
are analyzed for one-frequency pump beam. Followed that, in the case
of two-frequency pump, we find the backward first order scattering
mode is suppressed at the resonant condition $\Delta \omega =
8\omega_{r}$ and the forward second order mode is enhanced,
resulting from the combination of mode competition effects and
spatial distribution of the modes. The case of the three-frequency
pump beams for a lager number of the forward third order scattering
modes, and the higher modes for more resonant frequencies are
studied, which supplies a new method to get a large number of atoms
in higher order forward modes. Finally, some discussion and
conclusion are given.

\begin{figure}[tbp]
    \begin{center}
        \includegraphics[width=8cm]{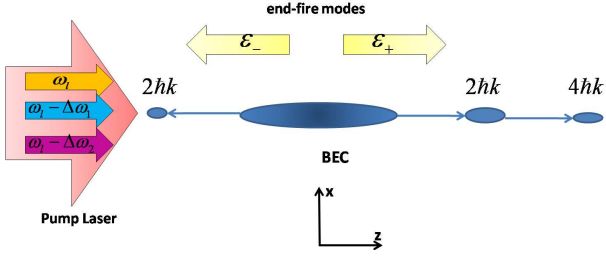}
    \end{center}
  \caption{(Color online) Our experimental scheme. A cigar-shape BEC
is illuminated by a far off-resonant laser pulse along its long axis
$\mathbf{\hat{z}}$. Collective Rayleigh scattering induces
superradiance. Two end-fire modes, which are also along
$\mathbf{\hat{z}}$ axis, form in superradiance process and the
1st-order recoiled atoms obtain a momentum of $2\hbar\mathbf{k}$.  }
\label{fig:1}
\end{figure}

\section{Model for a multiple-frequency end-pumped beam}
We consider the pump laser, with amplitude $\mathcal {E}_l(t)$,
polarization $\mathbf{e_y}$, wave vector $\mathbf{k_l}$, frequencies
$\omega_l$ and $\omega_{l}-\Delta \omega_n$, propagating along the
long axis $\mathbf{\hat{z}}$ of an elongated BEC,
$\mathbf{E}_l=\mathcal {E}_l(t)\mathbf{e_y}[(1+\Sigma_n e^{i\Delta
\omega_n t}) e^{i(k_lz-\omega_lt)}+c.c.]/2$,  as shown in Fig.\,1.
When supperradiant Rayleigh scattering happens, end-fire modes
spread along the same axis. The $\mathcal {E}_+$ mode has the same
direction as the incident light and mainly interacts with the right
part of the condensate, and the $\mathcal {E}_-$ mode overlaps with
the left part of the condensate. The atoms are recoiled to some
discrete momentum states with momentum $2m\hbar \mathbf{k}$, where
$m$ is an integer and the wave vector of end-fire mode light $k$ is
approximated as $ k_l$ for energy conservation. The total electric
field $\mathbf{E}(\mathbf{r},t)=\mathbf{E}^{(+)}+\mathbf{E}^{(-)}$
is given by~\cite{Bar-Gill2007arxiv,Zobay2006pra,yang,li}
\begin{eqnarray}
\mathbf{E}^{(+)}(\mathbf{r},t)&=&
 [(1 + \sum_n e^{i \Delta\omega_n t })
 \mathcal{E}_l(t) e^{-\mathrm{i}(\omega_lt-k_lz)}/2\nonumber\\
 &+& \mathcal {E}_-(z,t) e^{-\mathrm{i}(\omega t+kz)}]\mathbf{e_y}
\label{4}
\end{eqnarray}
where $\omega = ck$, $\mathbf{E}^{(-)}=\mathbf{E}^{(+)*}$, and
$\mathcal {E}_+$ is ignored because it has the same wave vector as
the pump beam but is very small in comparison to $\mathcal {E}_l$.
$\Delta \omega_n$ satisfies the condition $\Delta \omega_n \ll
\omega_l$~\cite{Bar-Gill2007arxiv} and the initial phases of the
different frequency components are assumed to be zero.

Since the BEC is tightly constrained in its short axis
 ($\mathbf{\hat{x}},\ \mathbf{\hat{y}}$) in the present
superradiance setup and the Fresnel number of the optical field is
around 1, one dimensional approximation is usually used
~\cite{Inouye1999science,Bar-Gill2007arxiv,li,Hilliard}. We expand
the wavefunction of the condensate $\psi(\mathbf{r},t)$ in momentum space, $\psi(\mathbf{r},t)=\sum_{m}
{\phi_m(z,t)}e^{-i(\omega_mt-2mkz)}$, where
$\phi_m(z,t)=\psi_m(z,t)/\sqrt{A}$, $\omega_m = 2\hbar m^2k^2/M$,
$m=0$ corresponds to the residual condensates, $m\neq0$ denotes the
side modes, and $A$ is the average cross area of the condensate
perpendicular to $\mathbf{\hat{z}}$.  Using the
Maxwell-Schr\"{o}dinger equations, we obtain dynamics
equations for $\phi_m(z,t)$,
\begin{eqnarray}
\mathrm{i}\frac{\partial \phi_m}{\partial
t}&=&-\frac{\hbar}{2M}\frac{\partial^2 \phi_{m}}{\partial z^2}
-\frac{2\mathrm{i}m\hbar k}{M}\frac{\partial \phi_m}{\partial z}
\nonumber\\
&+&\bar{g}\left[\mathcal{E}_-^*\phi_{m-1}
e^{-4\mathrm{i}(1-2m)\omega_rt}  +
\mathcal{E}_-\phi_{m+1}e^{-4\mathrm{i}(1+2m)\omega_rt}\right],\label{5}
\end{eqnarray}
where $\omega_{r} = \hbar k_{l}^2 / 2M$ is the recoil frequency, the
coupling between modes is given by
\begin{equation}
    \bar{g}(t) = g \left(1 + \sum_n e^{i \Delta\omega_n t } \right),
\end{equation}
with the coupling factor $g=\sqrt{3 \pi c^{3} R / (2 \omega_{l}^{2}
A L)}$,  $R$ is the Rayleigh scattering rate of the pump components,
and $L$ is the BEC length.

The first term on the right-hand-side of Eq.(\ref{5}) describes the
dispersion of $\phi_m$, and the second term gives rise to their
translation.  The terms in square brackets describe the atom
exchange between $\phi_m$ and $\phi_{m+1}$ or $\phi_{m-1}$
through the pump laser and end-fire mode fields. An atom in mode $m$
may absorb a laser photon and emit it into end-fire mode $\mathcal
{E}_-$, and the accompanying recoil drives the atom into $m+1$ mode,
hence atoms with mode $m+1$ can emerge in forward scattering. On
the other hand, in the backward scattering, atoms with mode $m$
absorb one $\mathcal {E}_-$ mode photon, deposit it into the laser
mode and go into mode $m-1$. The envelope function of end-fire
mode $\mathcal{E}_-$ is given by
\begin{equation}
\mathcal{E}_- = -\mathrm{i}\frac{\omega_r
\bar{g}}{2c\varepsilon_0}\int^{+\infty}_z\mathrm{d}z'
\sum_m\phi_m(z',t) \phi_{m+1}^*(z',t) e^{\mathrm{i}4(2m+1)\omega_{r}
t},\label{6}
\end{equation}
indicating that the end-fire mode field $\mathcal{E}_-$ is due to
the transition between $m$ and $m+1$ mode and the magnitude of
$\mathcal{E}_-$ depends on the spatial overlap between the two
modes.  In addition, there is a frequency difference of
$8\omega_{r}$ between adjacent modes.

\section{ The spatial and time evolution of scattered modes with a single-frequency
pump beam}

\begin{figure}[tbp]
\begin{center}
\includegraphics[height=4cm,width=6cm]{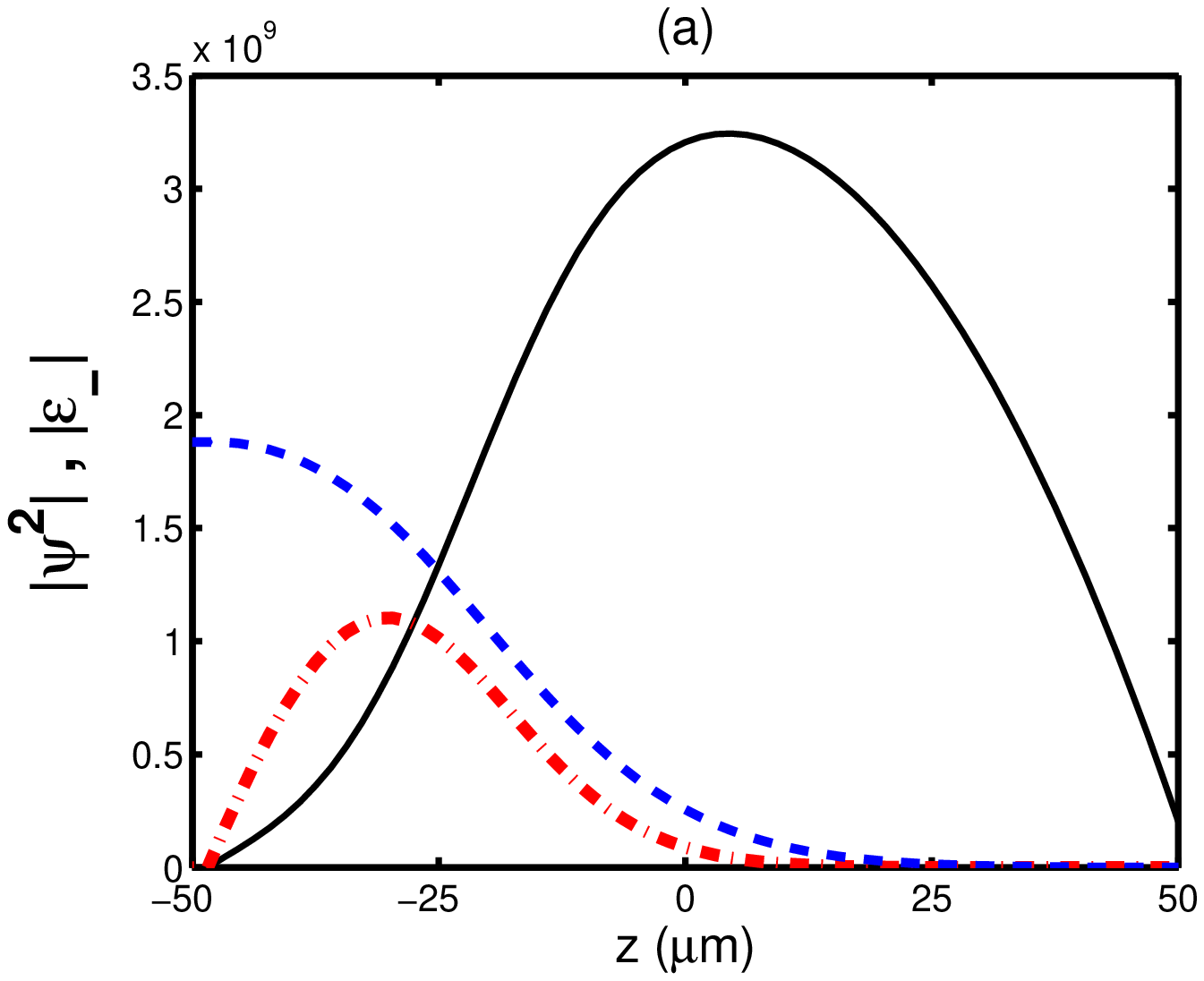}
\includegraphics[height=4cm,width=6cm]{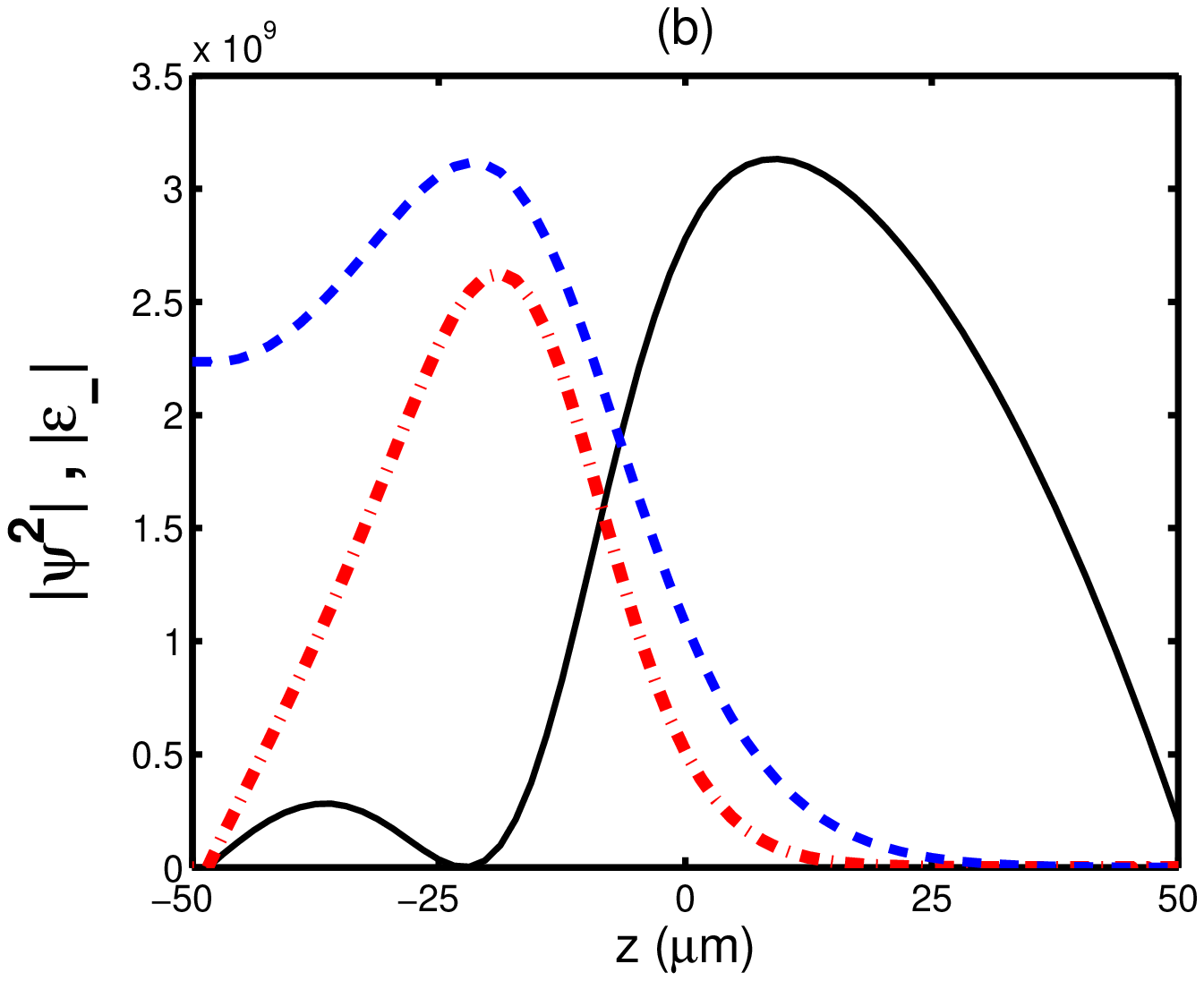}
\includegraphics[height=4cm,width=6cm]{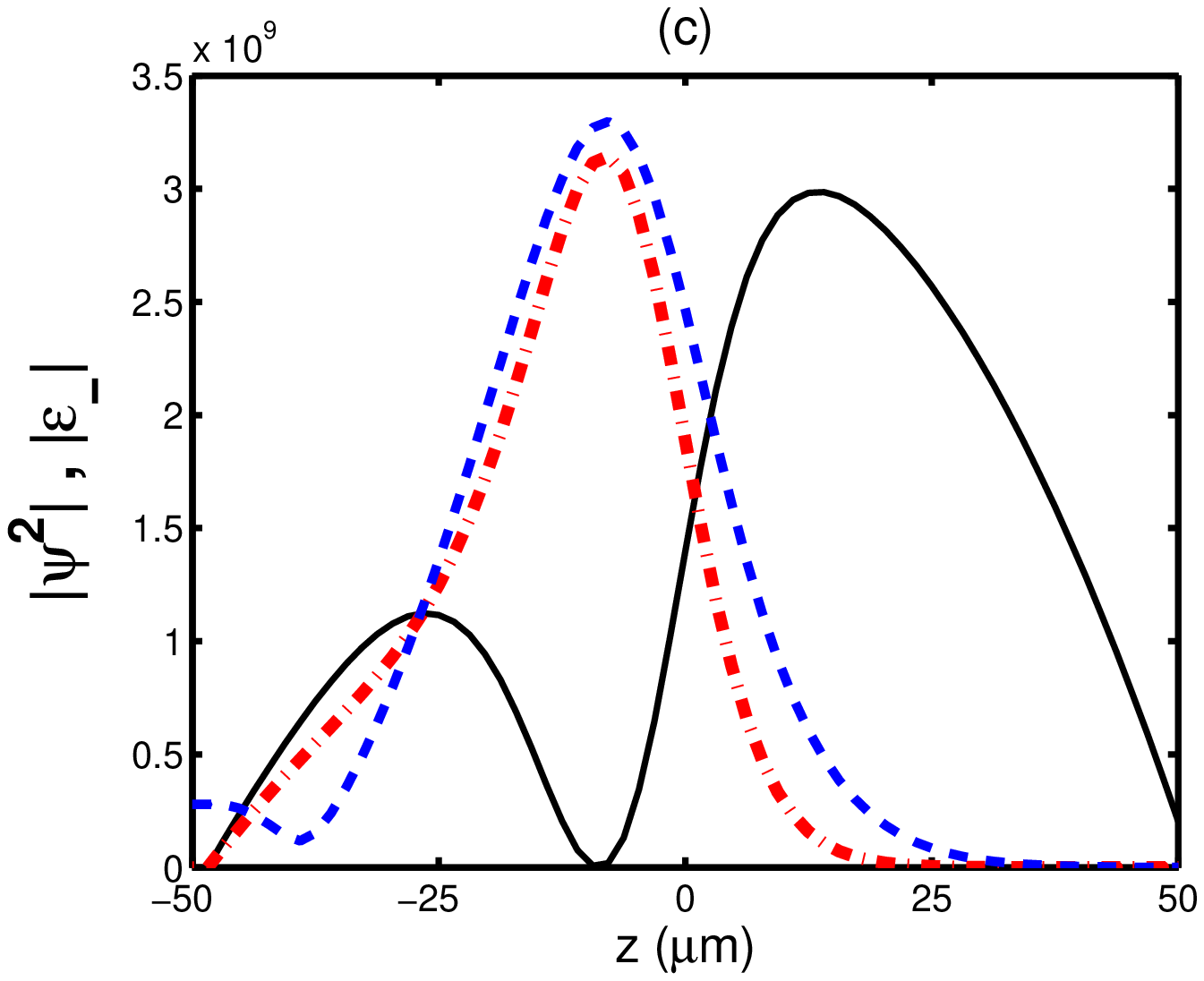}
\includegraphics[height=4cm,width=6cm]{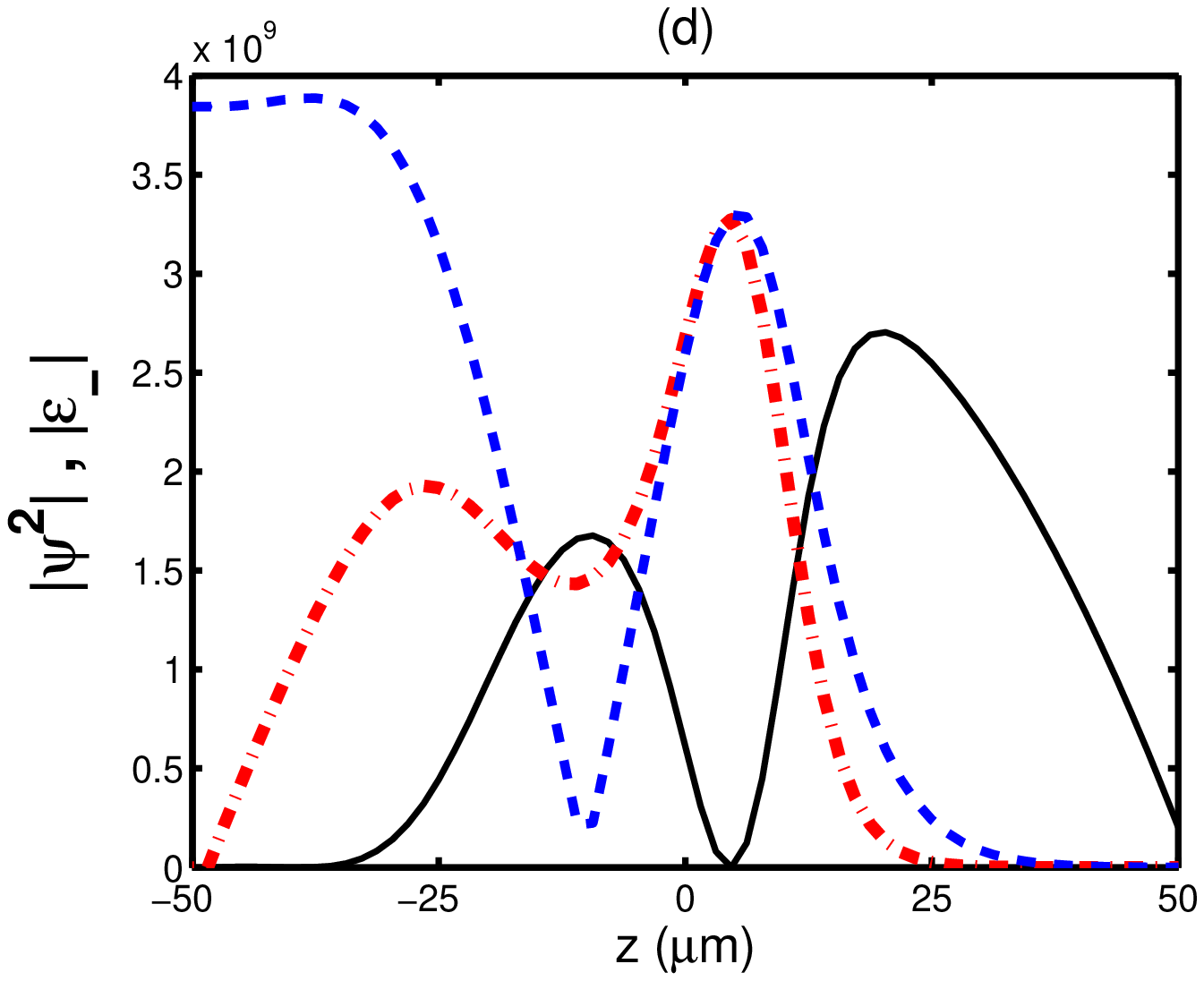}
\end{center}
\caption{(Color online) Spatial distribution of the atomic side
modes $|\psi^2|$ and the light end-fire mode $|\varepsilon_-|$ in
the weak coupling $(g=1.25\times10^6s^-1)$ in case of a single-frequency
pump for different pulse durations:  150us (a); 200us (b); 250us (c);
350us (d). The condensate mode $m=0$ is the solid line, the forward
first-order side mode $m=1$ is the dash-dotted line, and the
end-fire mode is the dashed line.} \label{fig:2}
\end{figure}

For explaining effects of spatial distribution and the depletion
mechanism in the scattering from BEC, we first study the case of a
single-frequency pump in the weak coupling regime.  The evolution of
spatial distribution of atomic side modes and optical end-fire mode
are depicted in Fig.\ref{fig:2}, where the original BEC is assumed
to be symmetrical.

Superradiance first starts at the leading-edge of the BEC, as
shown in Fig.\ref{fig:2} (a). The end-fire mode $\mathcal{E}_-$ monotonically
increases at the beginning and becomes strong on the side of
the end-pump, and it has a large overlap with the BEC. The
atomic side modes and the optical-field modes are well localized at
the condensate edge. Hence, the recoiled atoms mainly come from this
edge of the condensate, and the forward first order mode $m=1$
emerges due to the overlap between the condensate at $m=0$ and the
end fire mode $\mathcal{E}_-$.  Then at some point the condensate is
completely scattered to mode $m=1$ and the atoms are transferred
back to the edge, leading to a minimum in the condensate density and
regrowth at the edge, as shown in Fig.\ref{fig:2} (b), which appears
like a Rabi oscillation between the condensate and first-order side
mode. When the overlap between mode $m=0$ and $\mathcal{E}_-$ is
significantly large, the minimum point of mode $m=0$ and the peak of
mode $m=1$ move from the leading-edge to the center of the BEC, as
shown in Fig.\ref{fig:2} (c). When the regrowth part of mode $m=0$
is comparable to mode $m=1$, it will be scattered to mode $m=1$
again. Hence mode $m=1$ also has an edge regrowth.  Due to the
movement of the first peak and the regrowth from the edge, mode
$m=1$ will have a minimum point too, as shown in Fig.\ref{fig:2}
(d). This distribution shows the evolution of side modes in space
and the absence of backward-scattering modes in the weak
coupling regime.

The distribution of the first-order side mode closely connects with
the end-fire mode, and $\mathcal{E}_-$ is simply the result of the
coupling between $\phi_0$ and $\phi_1$. When the condensate
population at some point z is completely pumped to the first-order
side mode, the population of mode $m=1$ and $\mathcal{E}_-$ are at
maximum. When the first-order side mode absorbs end-fire mode
photons leading to the regrowth of the condensate, the populations of
mode $m=1$  and $\mathcal{E}_-$  will reach minimum. The asymmetry
could be explained by Eq.(\ref{5}), where  $\phi_0$ and $\phi_1$ are
coupled through $\mathcal{E}_-$ which is very small at the
tailing-edge of the condensate.

The evolution of the side modes and the end-fire mode indicates that
the scattering is a localized process. In this end-pumping
configuration, the scattering first starts on the leading edge of
the BEC and then moves towards the tailing edge.

\section{Mode competition for a two-frequency pump beam}

\begin{figure}[tbp]
\begin{center}
\includegraphics[width=8cm]{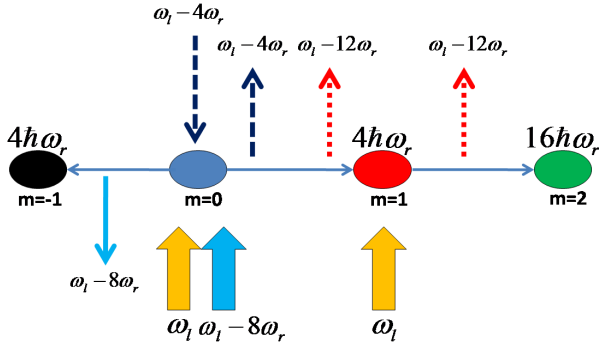}
\end{center}
\caption{(Color online) Light-field components of a
two-frequency pump laser. The broad arrows are the pump laser and
narrow ones are the end-fire mode (scattering optical field). In a
spontaneous process, atoms in the condensate absorb photons from the
pump laser with frequencies $\omega_{l}$ and
$\omega_{l}-8\omega_{r}$, are scattered to side mode $m=1$ and emit
end-fire mode photons with frequency $\omega_{l}-4\omega_{r}$
(dashed arrow) and $\omega_{l}-12\omega_{r}$ (dotted arrow),
respectively. Meanwhile, atoms in the condensate can also absorb
end-fire mode photons with frequency $\omega_{l}-4\omega_{r}$,  be
scattered back to side mode $m=-1$ and emit photons with frequency
$\omega_{l}-8\omega_{r}$(solid arrow), resonating to one of the pump
laser components.  The side mode $m=1$ can absorb pump laser photons
with frequency $\omega_{l}$ and be scattered to mode $m=2$, emitting
photons with frequency $\omega_{l}-12\omega_{r}$ resonating to the
existing end-fire mode.} \label{fig:comp}
\end{figure}

To investigate the effect of the two-frequency pump beam in the case
of end-pumping, the different frequency components of the end-fire
mode which indicate the energy change during the scattering are
depicted in Fig.~\ref{fig:comp}. The momentum of side mode $m=n$ is
$2n\hbar\textbf{k}$, and its kinetic energy is
$4n^2\hbar^2\textbf{k}^2/2M=4n^2\hbar\omega_{r}$. For the pump
component with frequency $\omega_{l}$, atoms from the condensate are
pumped to the side mode $m=1$ and emit end-fire mode photons with
frequency $\omega_{l}-4\omega_{r}$ spontaneously. However, in the
backward scattering process, an atom in the condensate absorbs the
end-fire mode ($\omega_{l}-4\omega_{r}$) and emits a photon with
frequency $\omega_{l}$ back into the pump laser. Since energy is not
conserved in backward-scattering, the backward side mode is not
populated in weak-pulse regime. Side mode $m=2$ is also not
populated due to the energy barrier. However, if we use the two
components pump laser with frequency difference $8\omega_{r}$, i.e.
resonant frequency difference, the energy mismatch can be
compensated by the pump laser.

Although the resonant condition for the backward mode is satisfied,
it should be noticed that two scattering channels exist almost
simultaneously. One is atoms scattered from side mode $m=0$ to
$m=-1$ and the other is from $m=1$ to $m=2$, resulting in mode
competition. The transition from mode $m=1$ to $m=2$ requires
absorbtion of photons from pump laser, while the backward transition
takes photons from the end-fire mode. Because the intensity of the
pump laser is far greater than that of the end-fire mode, the
transition from $m=1$ to $m=2$ has a bigger probability than the
transition from $m=0$ to $m=-1$.  Thus the population of the
backward mode $m=-1$ is suppressed even at the resonant condition,
while the forward mode $m=2$ is enhanced.

However, the existence of competition between these two channels may
not lead to the suppression of the backward mode. If these two
channels happen in different spacial parts of the condensate, then
both of side mode $m=-1$ and $m=2$ will be enhanced. The suppression
of backward mode $m=-1$ and the enhancement of mode $m=2$ need that
these two scattering channels happen in the same area. Therefore,
the spatial distribution effect should be considered.

\begin{figure}[tbp]
\begin{center}
\includegraphics[height=4cm,width=6cm]{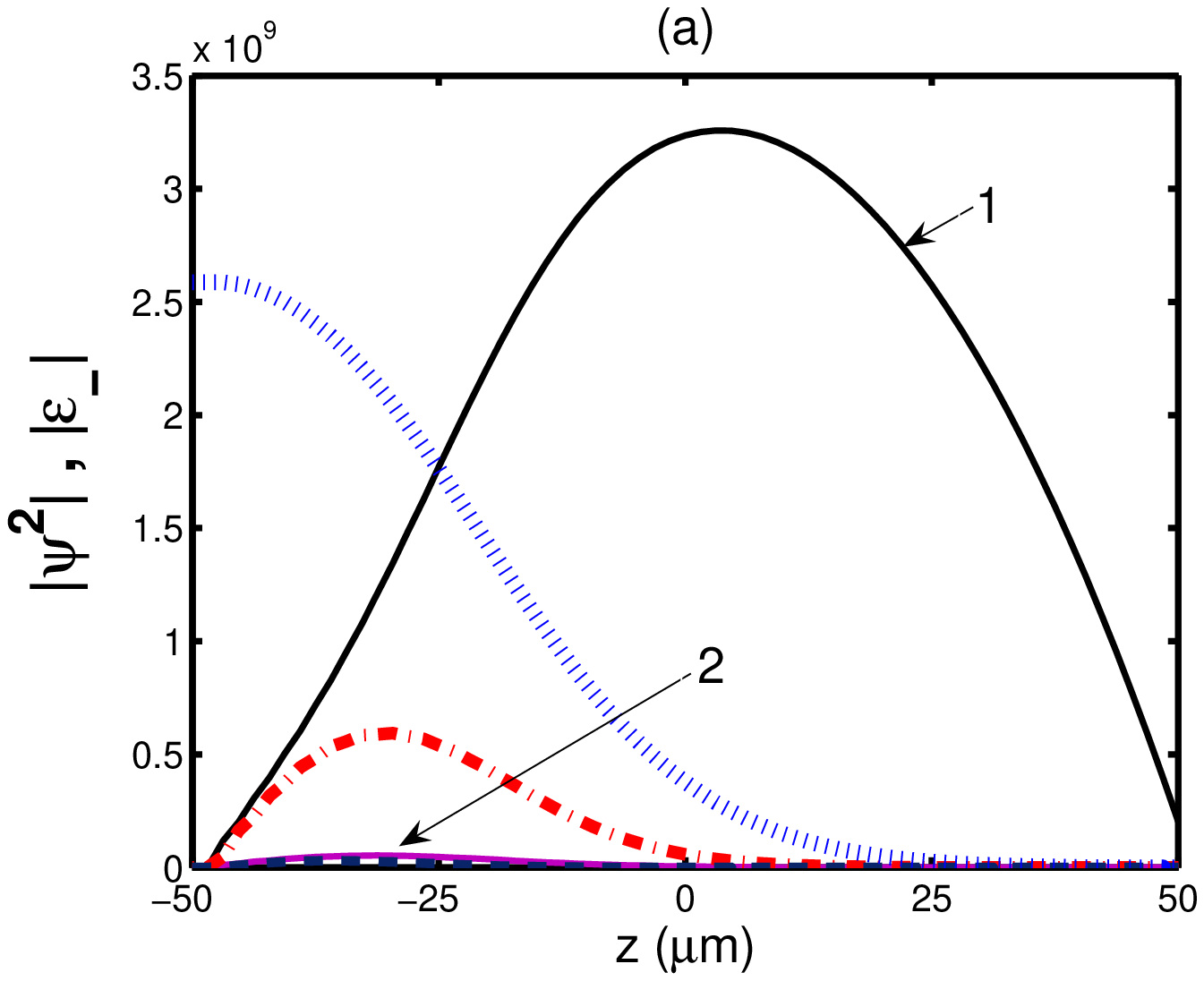}
\includegraphics[height=4cm,width=6cm]{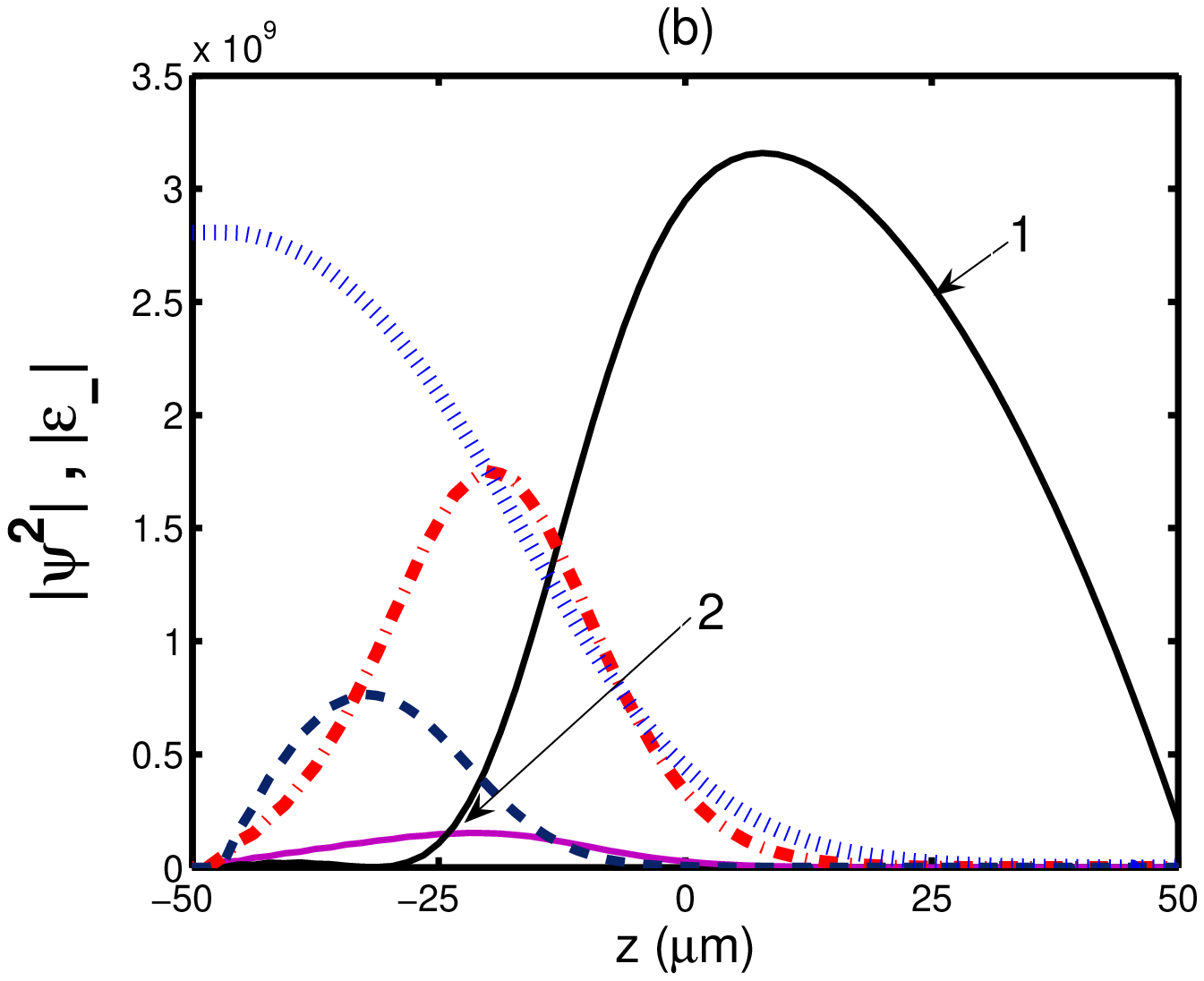}
\includegraphics[height=4cm,width=6cm]{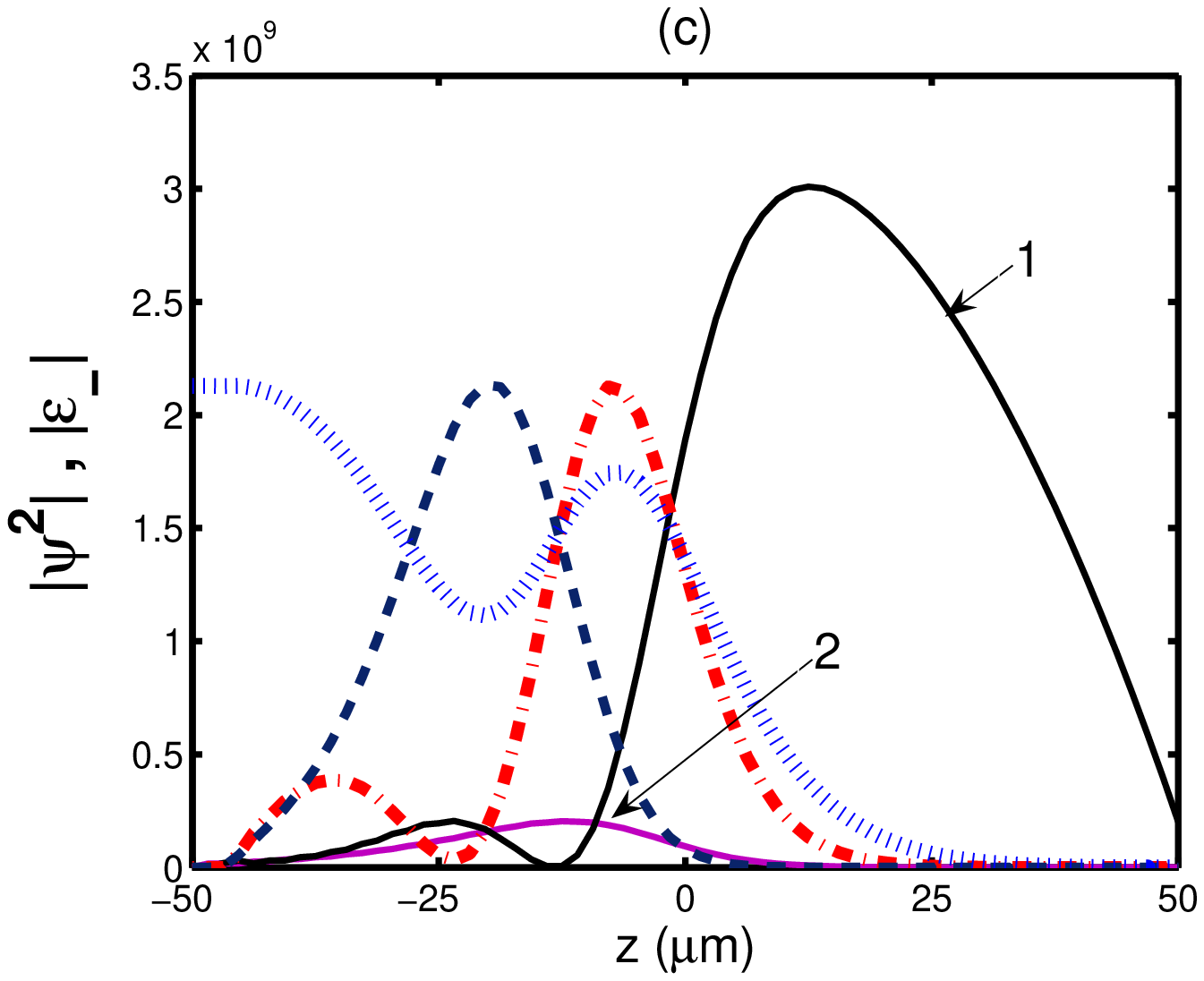}
\includegraphics[height=4cm,width=6cm]{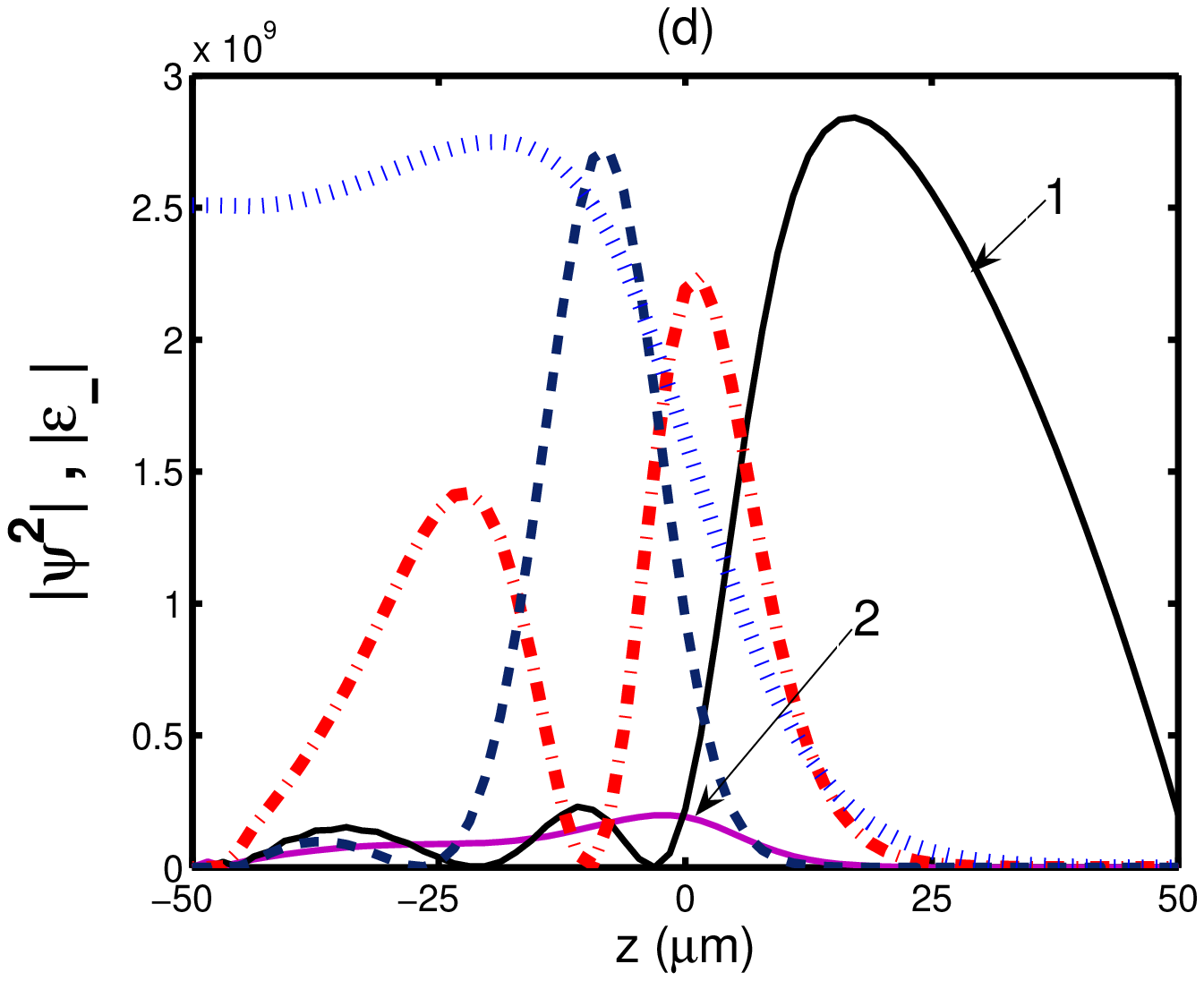}
\end{center}
\caption{(Color online) Spatial distribution of the side modes
$|\psi^2|$ and the end-fire mode $|\varepsilon_-|$ in the weak
coupling regime $(g=1.25\times10^6s^-1)$ with the two-frequency pump for
different pulse durations:  150us (a); 200us (b); 250us (c); 300us (d).
Condensate mode $m=0$ is the solid line-1, backward
first-order side mode $m=-1$ is the solid line-2, forward
first-order side mode $m=1$ is the dash-dotted line, forward
second-order side mode $m=2$ is the dashed line, and end-fire mode
is the dotted line.} \label{twofre}
\end{figure}
We analyze the spatial effect when second-order forward side
mode and backward side mode are populated at the resonant condition
$\Delta\omega=8\omega_{r}$.  The evolution of spatial distribution
of side modes and end-fire mode is shown in Fig.\ref{twofre}. Superradiance
first starts on the leading edge of the BEC, as shown
in Fig.\ref{twofre}(a). Although the backward first-order side mode
$m=-1$ is populated through the overlap between end-fire mode
$\mathcal{E}_-$ and side mode $m=0$, it is very small and emerges at
the leading-edge of the BEC.  Since the overlap between end-fire
mode and side mode $m=1$ is in the same area, the population of side
mode $m=2$ is obvious on this edge, as shown in Fig.\ref{twofre}(b).
Side mode $m=2$ grows more rapidly than side mode $m=-1$, which
means more atoms are scattered from side mode $m=1$ to $m=2$ than
that from $m=0$ to $m=-1$.Then the first peaks of side modes $m=1$
and $m=2$ move to the center of the BEC, as shown in
Fig.\ref{twofre}(c). Though the movement of the peaks is similar to
that in the case of a single-frequency pump laser, one major
difference is that the regrowth of side mode $m=0$ is very little,
hence nearly all the atoms on this edge are forwardly scattered. Due
to the nearly-complete depletion of the condensate, atoms are mainly
transferred between side mode $m=1$ and $m=2$. The apparent regrowth
of side mode $m=1$ on the leading-edge shown in Fig.\ref{twofre}(d)
indicates that there are Rabi oscillations between side modes
$m=1$ and $m=2$ in the depleted area of the condensate.

The above phenomenon is different from the case of the pump laser
traveling along the short axis. In the latter case, a correlation
between the center depletion of the BEC and backward mode was
reported in Ref.~\cite{Zobay2006pra}. However, in our case, such
correlation does not exist because side mode $m=2$ emerges on the
edge of the BEC. As a consequence of the edge depletion of the BEC,
backward side mode $m=-1$ is not populated significantly, because
the end-fire mode mainly distributes in the leading edge where side
mode $m=0$ suffers the strongest depletion.  Thus only a small
number of atoms in the residual condensate can absorb end-fire mode
photons and be scattered backwardly. In another word, the emergence
of side mode $m=2$ suppress backward-scattering atoms. Therefore the
efficiency of getting  $m=2$ mode with this two-frequency pump is
strongly enhanced while greatly suppressed for the backward side
mode.

\begin{figure}[tbp]
\begin{center}
\includegraphics[height=4cm,width=6cm]{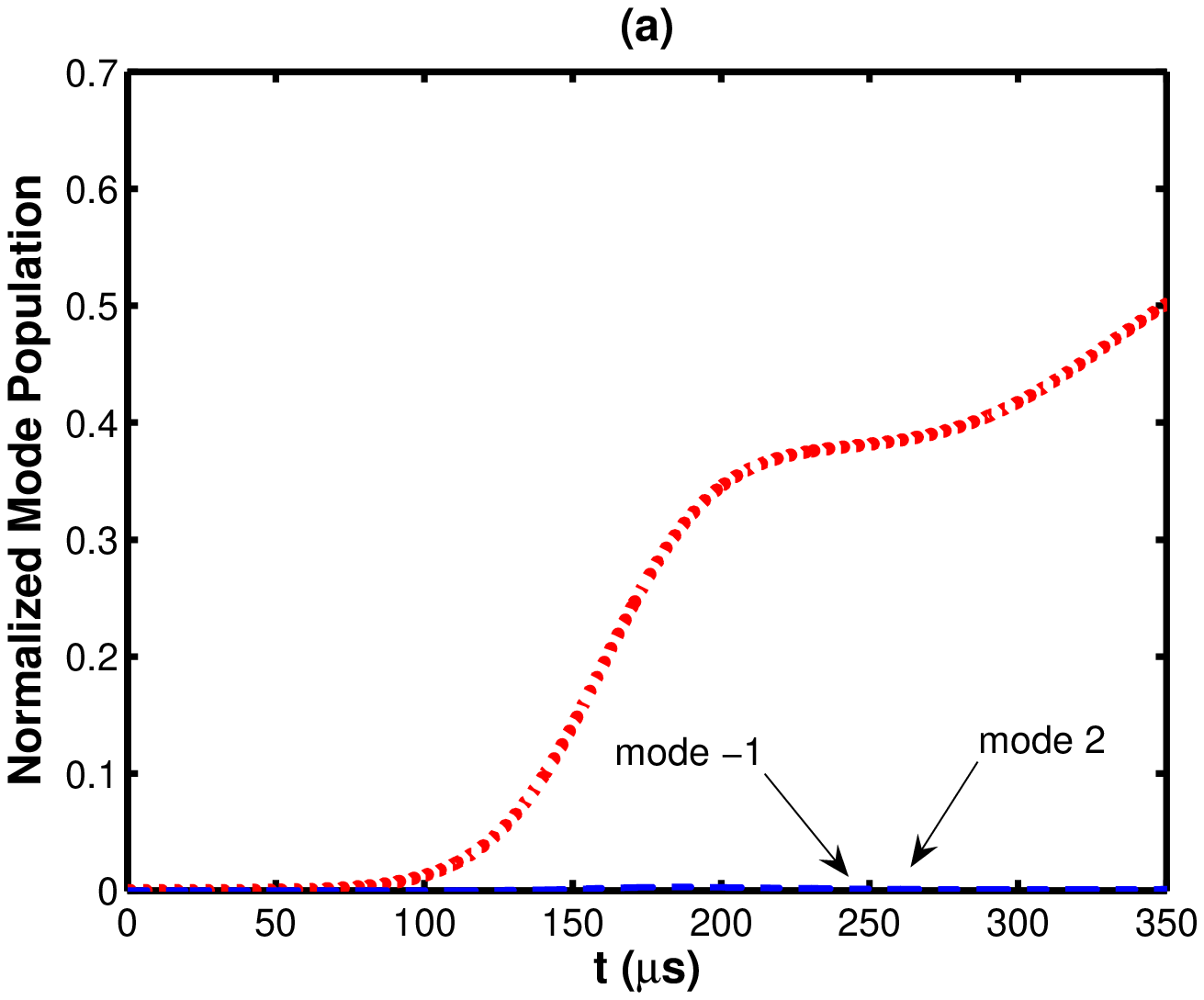}
\includegraphics[height=4cm,width=6cm]{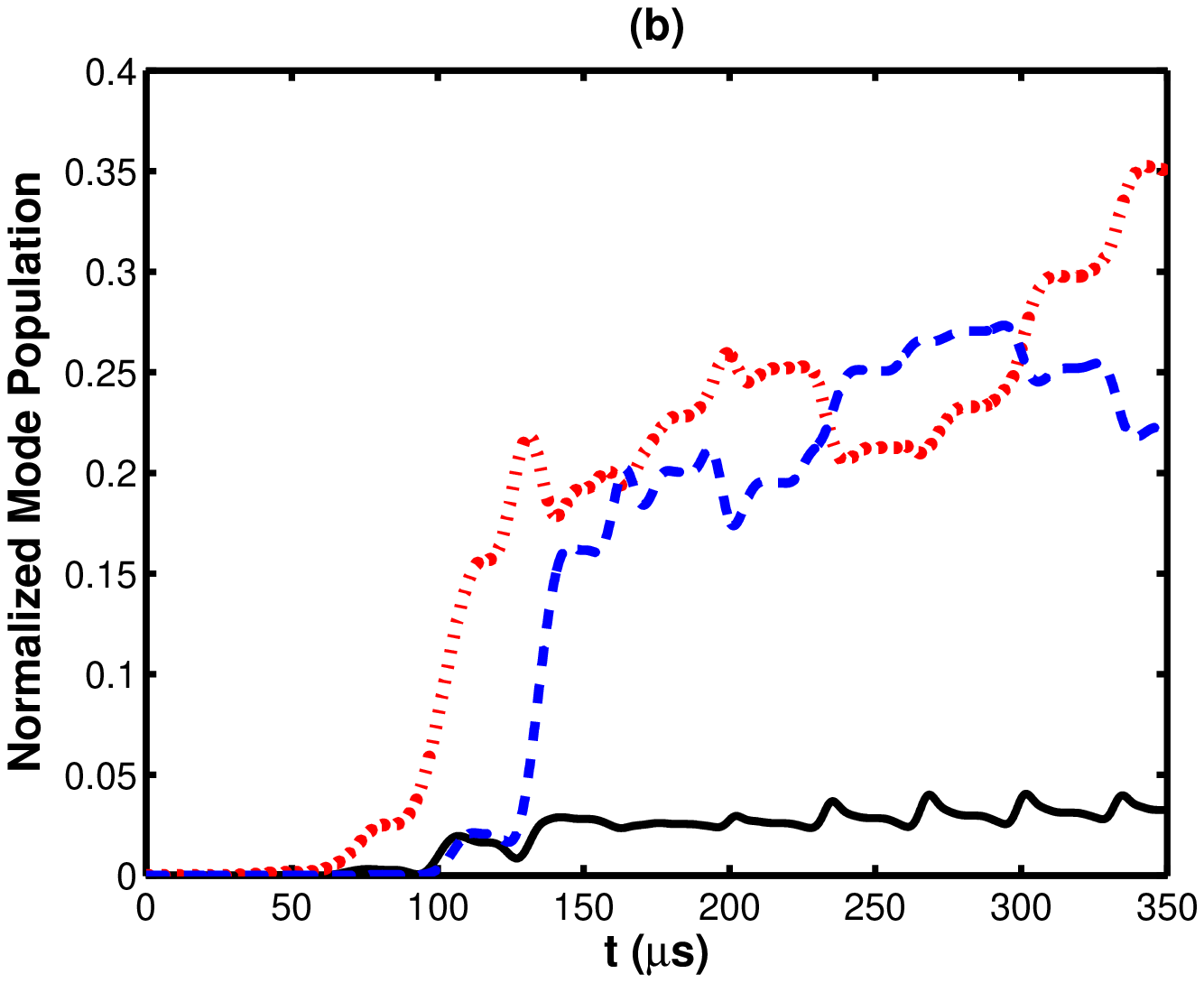}
\end{center}
\caption{(Color online) Normalized side mode populations versus
time: (a) for a single-frequency pump beam; (b) for a two-frequency
resonant pump beam. In both cases the coupling constant is kept
$g=1.55\times 10^6 s^{-1}$. The side mode are: m=-1 (solid); m=1
(dotted); m=2 (dashed). } \label{fig:1vs2}
\end{figure}

The time evolution of several side modes populations normalized by
the total atom number are depicted by Fig.\ref{fig:1vs2}.
Fig.\ref{fig:1vs2} (a) shows that using a single-frequency pump
laser cannot produce backward mode $m=-1$ or forward higher mode
$m=2$ in the weak-pulse regime. Using a resonant two-frequency pump
beam with the same intensity, modes $m=-1$ and $m=2$ increased, as
shown in Fig.\ref{fig:1vs2} (b), however, the forward mode is
greatly enhanced while the backward mode remains very small.

\section{The third order forward modes Enhanced with a three-frequency pump beam}
\begin{figure}[tbp]
\begin{center}
\includegraphics[width=8cm]{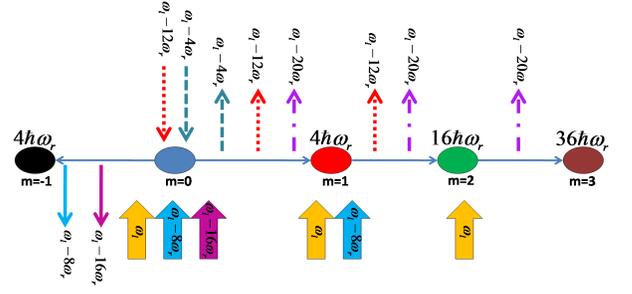}
\end{center}
\caption{(Color online) The light-field components of a
three-frequency pump laser. The broad arrows are the pump laser and
narrow ones are the end-fire mode. } \label{fig:comp3}
\end{figure}
The second forward side mode $m=2$ is greatly enhanced with a
resonant two-frequency pump beam, however, the populations of higher
forward modes such as $m=3$ are very small as the channel from the
second forward mode to the third forward mode is not resonant with
the exiting optical field. To get a large number mode for $m=3$,
Fig.\ref{fig:comp3} depicts the scheme of the three-frequency pump
beam with the frequencies of the pump laser $\omega_l$,
$\omega_l-8\omega_r$ and $\omega_l-16\omega_r$. The frequency
components $\omega_l$, $\omega_l-8\omega_r$ and
$\omega_l-16\omega_r$ both have the resonant frequency difference.
Hence, there could be two channels to form the backward side mode
$m=-1$ but the enhancement of the backward scattering is small
because of the formation of higher forward side modes. There are
also two channels to form side mode $m=2$. One thing different from
the two-frequency pump beam is that there is also a channel to form
side mode $m=3$ for the reason that atoms in side mode $m=2$ absorb
pump laser photons with frequency $\omega_l$, are then scattered to
mode $m=3$ and eventually emit end-fire mode photons with frequency
$\omega_l-20\omega_r$ which is resonant to an existing end-fire
mode. This means that more atoms in side mode $m=2$ will be pumped
to side mode $m=3$ and less will be transferred back to side mode
$m=1$, a competition between side mode $m=3$ and $m=1$ is set up. As
a result, side mode $m=3$ will be enhanced and $m=1$ will be reduced
relatively.

\begin{figure}[tbp]
\begin{center}
\includegraphics[height=4cm,width=6cm]{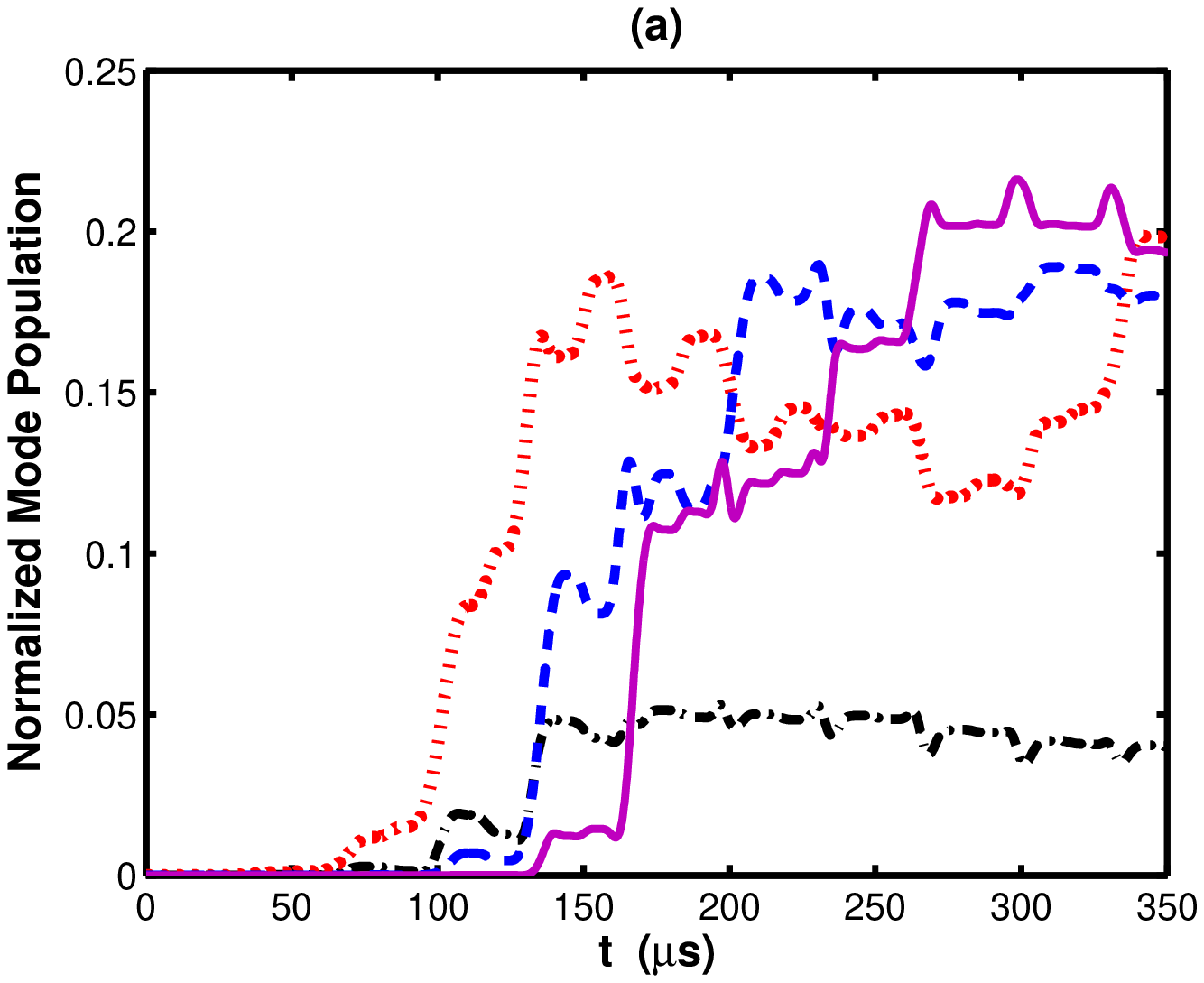}
\includegraphics[height=4cm,width=6cm]{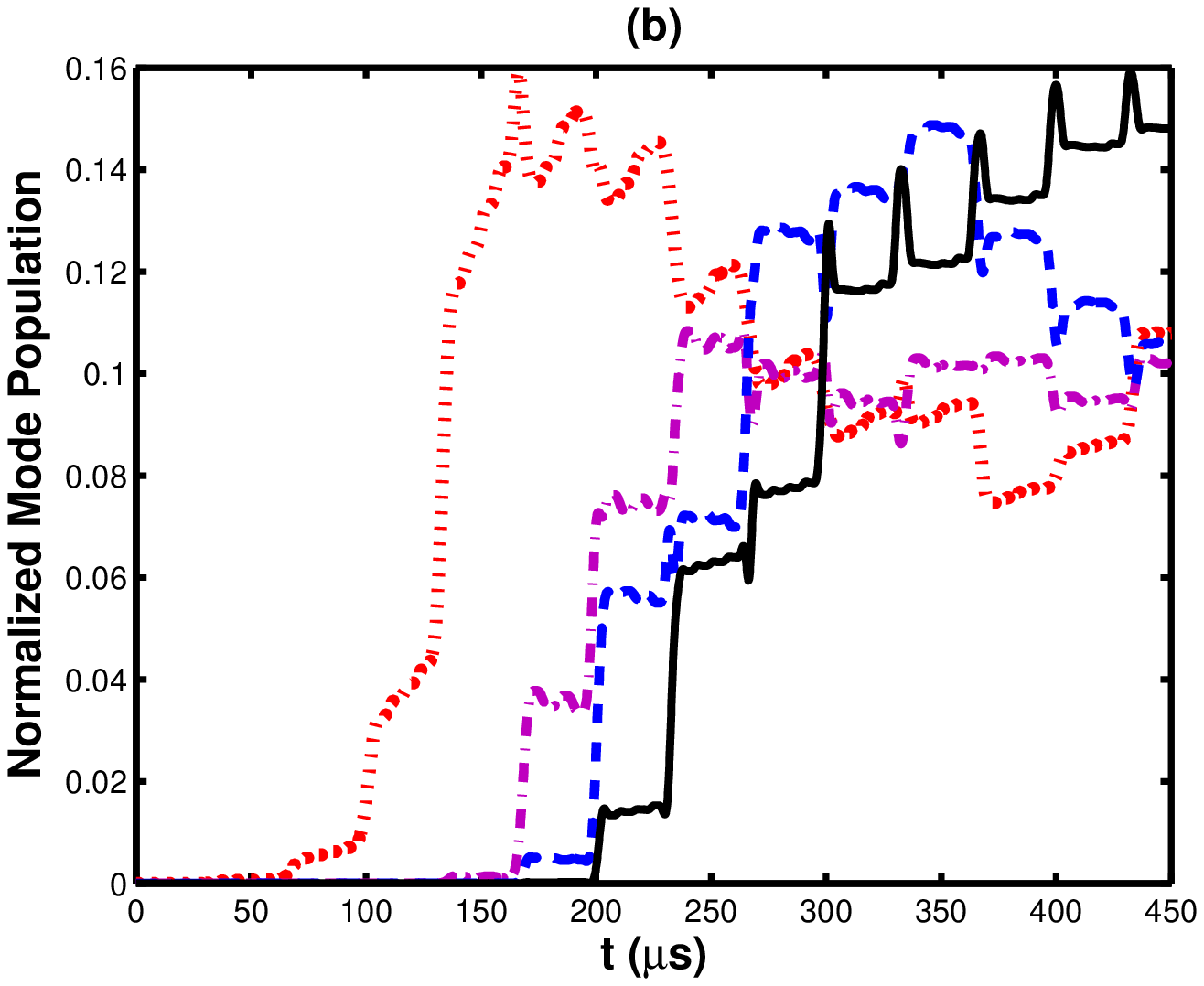}
\end{center}
\caption{(Color online) Normalized side mode populations versus time
with the coupling constant $g=1.55\times 10^6 s^{-1}$: (a) for a
three-frequency pump laser: $m=-1$ (dash-dotted), $m=1$ (dotted),
$m=2$ (dashed), $m=3$ (solid) ; (b) for a five-frequency pump laser:
$m=1$ (dotted), $m=3$ (dash-dotted), $m=4$ (dashed), $m=5$ (solid)
.} \label{fig:2vs3}
\end{figure}

Fig.\ref{fig:2vs3}(a) is the simulated result of the time evolution
of normalized side mode populations for a three-frequency pump beam.
We could see that side mode $m=3$ would be strongly enhanced at long
time while side mode $m=1$ reduced.

\section{Momentum transfer in the high order forward modes}

From the above discussion we know that using multi-resonant
frequencies is a promising way to get a large number of higher
forward modes. When a pump laser has frequency components $\omega_l,
\omega_l-8\omega_r, \cdots, \omega_l-(n-1)*8\omega_r$, satisfying
$(n-1)*8\omega_r\ll\omega_l$, with the kinetic energy of mode $m=n$
equal to
 $4n^2\hbar\omega_r$, then after the condensate atoms spontaneously
scattered to mode $m=1$, the end-fire mode will have frequency
components $\omega_l-4\omega_r, \omega_l-12\omega_r, \cdots,
\omega_l-(2n-1)*4\omega_r$. For resonance concern, mode $m=1$ will
absorb photons from the pump components $\omega_l,
\omega_l-8\omega_r, \cdots, \omega_l-(n-2)*8\omega_r$ and emits
end-fire mode photons with frequency $\omega_l-12\omega_r, \cdots,
\omega_l-(2n-1)*4\omega_r$ which are resonant with existing end-fire
mode, so mode $m=2$ is produced. Like mode $m=1$, modes $m=2, m=3,
\cdots, m=n-1$ can absorb pump photons and emit photons resonant to
the existing end-fire mode. For example, mode $m=n-1$ will absorb
photons with frequency $\omega_l$ and emits photons with frequency $
\omega_l-(2n-1)*4\omega_r$. Therefore atoms could finally be
transferred to mode $m=n$. Note that mode $m=n$ cannot emit resonant
end-fire mode, so mode $m=n$ will be enhanced. To show it,
Fig.\ref{fig:2vs3}(b) is the simulated result of the time evolution
of normalized side mode populations for a five-frequency pump beam.
We could see that side mode $m=5$ would be strongly enhanced.

\section{Discussion and Conclusion}

In the experiment, to get several resonant frequencies, the laser
beam from an external cavity diode laser can be split into several
parts, and their frequencies are shifted individually by
acoustic-optical modulators (AOMs) which are driven by phase-locked
radio frequency signals, as demonstrated in the case of two resonant
frequency~\cite{Bar-Gill2007arxiv,yang}. Therefore, the frequency
difference between the beams can be controlled precisely.
Furthermore, to avoid the reflection from the glass tube and
formation of Bragg scattering in the experiment, the pump beam can
actually deviate a few degrees from the long axis, as shown in the
experiments~\cite{2005Italy, li, Hilliard}.

Different to the works in the configuration where the pump beam
travels along the short axis of the condensate with the resonant
frequency~\cite{yang}, where a large number of backward scattering
is obvious in a two-frequency pump beam, the backward scattering is
suppressed and the forward second-order mode is obviously enhanced
in our case. This is due to mode competition between the forward
second-order mode and the backward mode and local depletion of the
superradiant process.

We have not considered the initial quantum process because its time
scale is very small, shorter than $1 \mu s$. In this quantum process
there is also mode competition to form the end-fire modes along the
long axis and suppress the emission on the other direction. This is
 different concept from what has been discussed above, in which case mode
competition exists in the different channels satisfying the energy
match and spatial condition.

For the pump beam with several resonant frequencies, not only can we
obtain the high order momentum transfer which is important in the
momentum manipulation for atom interferometry, but also the above
analysis is useful to understand the interplay between the matter
wave and light in the matter wave
amplification~\cite{Schneble2003scince,1999}, atomic cooperative
scattering in the optical lattice~\cite{xu},  and by the pump with a
noisy laser ~\cite{robb,zhou}.

In conclusion, superradiant scattering from BEC is studied with
incident light having different frequency components traveling along
the long axis of the BEC in the weak coupling regime. It provides a
method to get high forward modes by adding different frequency
components to the pump beam. This is the result of both mode
competition for the concern of energy and the local depletion of the
spatial distribution. Our results shows that the spatial effects and
mode competition are very important even in the case of resonant
superradiance.

We thank Thibault Vogt, Lan Yin for critical reading of the
manuscript and comments. Thank L. You for his helpful discussion.
This work is partially supported by the state Key Development
Program for Basic Research of China (No.2005CB724503, and
2006CB921402,921401), and by NSFC (No.10874008, 10934010 and
60490280).

\end{document}